# Description of the Grover algorithm based on geometric considerations

G. Fleury and P. Lacomme

*Université Clermont-Auvergne, Clermont-Auvergne-INP, LIMOS (UMR CNRS 6158),
1 rue de la Chebarde,
63178 Aubière Cedex, France*

gerard.fleury@isima.fr, philippe.lacomme@isima.fr

*Abstract*

This paper concerns the Grover algorithm that permits to make amplification of quantum states previously tagged by an Oracle. Grover's algorithm allows searches in an unstructured database of $n$ entries, finding a marked element with a quadratic speedup. The algorithm requires a predefined number of runs to succeed with probability close to one.

This article provides a description of the amplitude amplification quantum algorithm mechanism in a very short computational way, based on tensor products and provides a geometric presentation of the successive system states. All the basis changes are fully described to provide an alternative to the wide spread Grover description based only on matrices and complex tensor computation. Our experiments encompass numerical evaluations of circuit using the Qiskit library of IBM that meet the theoretical considerations.

## 1. Introduction

Quantum Computing received a considerable of interest from the physic community but has received less attention in the Operational Research (OR) community that could be surprising considering the potential of quantum computing in OR perspective.

A typical problem in Operational Research lies on the investigation into a landscape of solutions satisfying a set of conditions. Most of the time the difficulty comes from the very large number of solutions in the search space and second from the very small number of points that model a solution. For example in a Job-Shop composed of $n = 10$ jobs and $m = 10$ machines the total number of disjunctive graphs is larger than $10^{50}$ and only a very small number of these graphs are acyclic graphs that model a solution. However, as stressed by (Grover, 1996), the most famous problem in the OR community remains the SAT that consists in determining if it is possible or not to assign a value to a set of $n$ binary variables to satisfy a set of clauses $C$.

This paper focuses on one of the most important quantum algorithms introduced by Grover in 1996 that permits a simultaneous evaluation of all the SAT problems to find the correct assignment with a promise of a quadratic speedup.

Most of the papers in the current literature that introduce the Grover's algorithm comes from physic journal and our objective remains in introducing the algorithm to the OR community and to illustrate the Grover's principles using geometric considerations and to give a very compact and readable tensorial computation of the successive quantum states.

## 2. Grover approach

*2.1. Principles and notations*

Let us consider an unsorted finite set $B$ spanning a Hilbert space $E = Span(B)$ and a function

$$f: \quad B \to \{0; 1\}$$
$$x \to f(x)$$

referred to as an Oracle that characterizes the marked subsets of $B$ with

$$E_1 = span(x \in B / f(x) = i)$$

and

$$E = E_0 \oplus E_1$$

The problem consists in finding at least one $x \in E_1$ avoiding the costly enumeration of all element of $B$ one by one if no extra information is available on $B$. Because search procedures are the corner stone in computer science of advanced data structures, the Grover's algorithm that provides a quadratic speed-up received a considerable of attention.

An amplification procedure consists in considering first an initial $|\psi\rangle \in E$ to return a state $|\psi\rangle \in E_1$ with a probability close to 1. Note that multiple marked elements by the Oracle do not change the Grover's principles but the number of amplifications has to be tuned (Grover, 1996) (Zalka, 1999). The Grover amplification efficiency lies on an expected number of calls to the Oracle.

The Hadamard gate is an operation that maps the basis state $B(|0\rangle; |1\rangle)$ into $B(|p\rangle; |m\rangle)$ creating an equal superposition of the two basis states.

$H.|0\rangle = \dfrac{1}{\sqrt{2}} \cdot (|0\rangle + |1\rangle) = |p\rangle$

$H.|1\rangle = \dfrac{1}{\sqrt{2}} \cdot (|0\rangle - |1\rangle) = |m\rangle$

$H.|p\rangle = |0\rangle$

$H.|m\rangle = |1\rangle$

The X-gate is a symmetry around the $\pi/4$ axis leading to the following basic transformations.

$X.|0\rangle = |1\rangle$

$X.|1\rangle = |0\rangle$

$H.|p\rangle = |p\rangle$

$H.|m\rangle = -|m\rangle$

Note that the notations $|p\rangle$ for $H.|0\rangle$ and $|m\rangle$ for $H.|1\rangle$ are used for conveniences but the common notation is $|+\rangle$ and $|-\rangle$.

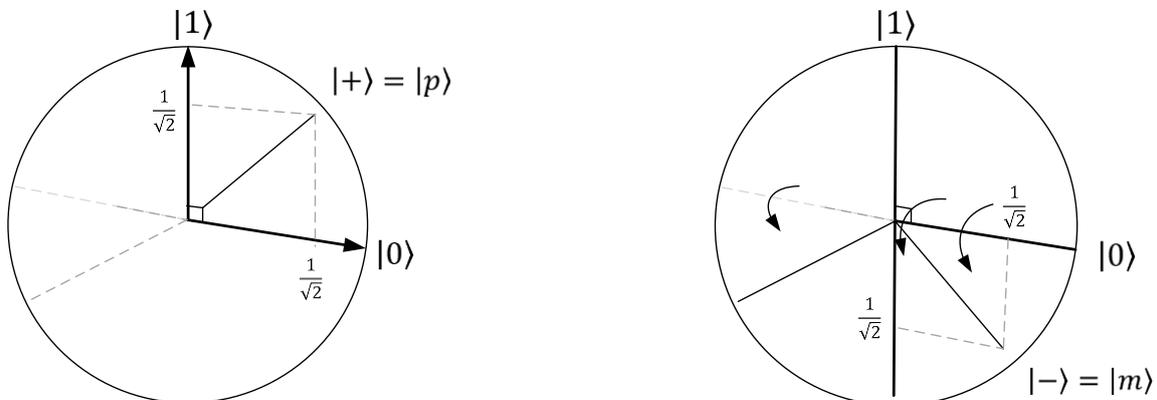

Figure1. Basic operation visualization

The minus sign is used to define the opposite direction of one state $|\psi\rangle$ as stressed on figure 1 where $-|1\rangle$ is the opposite direction to $|1\rangle$ meaning that the phase of $-|1\rangle$ is $\pi$.

*2.1 Grover geometric description for $r = |10100\rangle$*

Figure 2 gives the Grover circuit for $n = 5$ and $r = |10100\rangle$. We provide a geometric description where we alternate in computation with $n$ and the specific value 5 depending on formulae and comments on the figures.

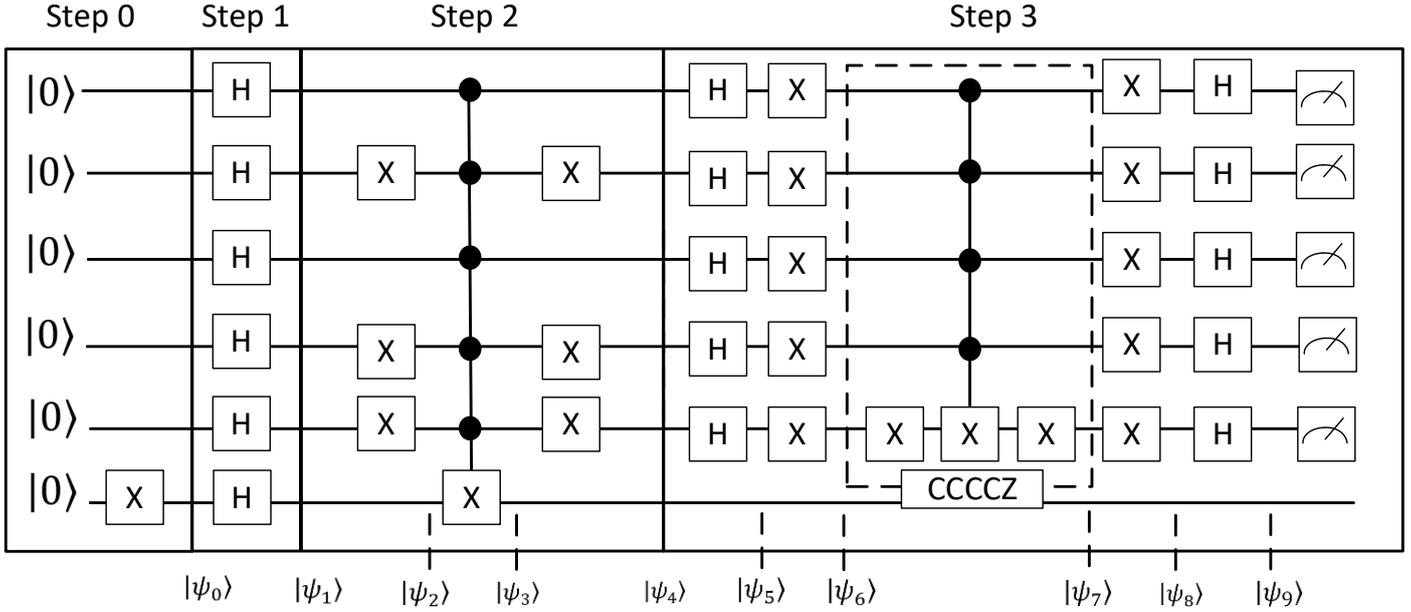

Figure 2. Grover's circuit for $n = 5$ and $r = |10100\rangle$

**Step 0 and 1.** Initialization

Step 0. $|\psi_0\rangle = |0\rangle^{\otimes 5} \otimes |1\rangle$

The qubit number 6 is required only for the Oracle, the 5 first ones are useful for the problem to solve.

Step 1. Application of $H^{\otimes 6}$ assigns a similar amplitude to all states with half of states with positive amplitude and half with negative one in the computational base $B(|0\rangle; |1\rangle)$.

$$|\psi_1\rangle = H^{\otimes 6}.|\psi_0\rangle$$
$$|\psi_1\rangle = H^{\otimes n}.|0\rangle^{\otimes 5} \otimes H.|1\rangle$$
$$|\psi_1\rangle = |p\rangle^{\otimes 5} \otimes |m\rangle$$

The operator $H^{\otimes 6}$ applied to $|\psi_0\rangle$ gives an equal superposition of the states that is a uniform superposition of all elements in the basis $B(|p\rangle^{\otimes 5}; |m\rangle)$. In the computational base with an amplitude of $\frac{1}{\sqrt{2^5}}$. The probability of each state is $\left(\frac{1}{\sqrt{2^5}}\right)^2 = \frac{1}{2^5}$.

In the basis $B(|p\rangle^{\otimes 5}; |m\rangle)$, $|p\rangle^{\otimes 5}$ is an eigenvector of $X^{\otimes n}$ assigned to the eigenvalue 1 and $|m\rangle$ is an eigenvector of $X$ assigned to $-1$ which is the second eigenvalue of $X$.

**Step 2.** Oracle definition

The Oracle consists in performing one $CC\ldots CX$ to change the phase of every $|r\rangle \in E_1$ by $\pi$. After applying $CC\ldots CX$ the quantum state (limited to the $n$ first qubits) should be:
$$|\psi\rangle = |p\rangle^{\otimes 5} - |10100\rangle$$

The **first step** of the Oracle consists in an application of $(Id \otimes X \otimes Id \otimes X \otimes X \otimes Id)$.

For $X.|p\rangle = |p\rangle$ we have:
$$|\psi_2\rangle = (Id \otimes X \otimes Id \otimes X \otimes X \otimes Id).(|p\rangle^{\otimes 5} \otimes |m\rangle)$$
$$|\psi_2\rangle = (Id \otimes X \otimes Id \otimes X \otimes X).(|p\rangle^{\otimes 5}) \otimes |m\rangle$$
$$|\psi_2\rangle = |p\rangle^{\otimes 5} \otimes |m\rangle$$

Hence in the plane spanned by $|r\rangle = |10100\rangle$ and $|p\rangle^{\otimes 5}$ we have:
$$(Id \otimes X \otimes Id \otimes X \otimes X \otimes Id).|10100\rangle \otimes |m\rangle = |1\rangle^{\otimes 5} \otimes |m\rangle$$
and
$$(Id \otimes X \otimes Id \otimes X \otimes X \otimes Id).|p\rangle^{\otimes n} \otimes |m\rangle = |p\rangle^{\otimes 5} \otimes |m\rangle$$

Considering only the 5-first qubits, this quantum state means that the base $B(|r\rangle; |p\rangle^{\otimes 5})$ has been switched into $B(|1\rangle^{\otimes 5}; |p\rangle^{\otimes 5})$

The **second step** consists in application of $CCCCCX$ that does not affect $B(|1\rangle^{\otimes 5}; |p\rangle^{\otimes 5})$ and only switches $|m\rangle$ in $-|m\rangle$. Note that the $CCCCCX$ is activated only in the base spanned with $|1\rangle^{\otimes 5}$.
We have
$$CCCCCX.(|1\rangle^{\otimes 5} \otimes |m\rangle) = |1\rangle^{\otimes 5} \otimes -|m\rangle = -|1\rangle^{\otimes 5} \otimes |m\rangle$$
and
$$CCCCCX.(|p\rangle^{\otimes 5} \otimes |m\rangle) = |p\rangle^{\otimes 5} \otimes |m\rangle - \frac{2}{\sqrt{2^n}}|1\rangle^{\otimes 5} \otimes |m\rangle$$
$$CCCCCX.(|p\rangle^{\otimes 5} \otimes |m\rangle) = \left(|p\rangle^{\otimes 5} - \frac{2}{\sqrt{2^n}}|1\rangle^{\otimes 5}\right) \otimes |m\rangle$$

For convenience, and without loss of generality and because the ansatz is not performed by any gate after the Oracle, every representations are only concerned by the first $n$ qubits. Note also it is possible to use $C..CZ - gate$ on the 5 first qubits or $C..CX$ on the 6 qubits for convenience.

We have:
$$|\psi_3\rangle = |p\rangle^{\otimes 5} - \frac{2}{\sqrt{2^n}}|1\rangle^{\otimes 5}$$

The **last step** of the Oracle consists in application of $(Id \otimes X \otimes Id \otimes X \otimes X)$ that switches:
$$(Id \otimes X \otimes Id \otimes X \otimes X).(-|1\rangle^{\otimes 5}) = -|10100\rangle$$
for $X.|1\rangle = |0\rangle$ and

$$(Id \otimes X \otimes Id \otimes X \otimes X).|p\rangle^{\otimes 5} = |p\rangle^{\otimes 5}$$
meaning that the base $B(|1\rangle^{\otimes 5}; |p\rangle^{\otimes 5})$ turns back into $B(-|10100\rangle; |p\rangle^{\otimes 5})$

In the very specific situation where $|r\rangle = |10100\rangle$, the geometric representation of figure 3, shows that the state after the oracle is in the plane spanned by $|p\rangle^{\otimes 5}$ and by $|r\rangle$.
$$|\psi_4\rangle = |p\rangle^{\otimes 5} - \frac{2}{\sqrt{2^n}}|10100\rangle$$

The state vector $|p\rangle^{\otimes 5}$ (before performing $CC \ldots CX$) is after application of the oracle $|\psi\rangle = |p\rangle^{\otimes 5} - \frac{2}{\sqrt{2^n}}.|r\rangle$:
- The state $|p\rangle^{\otimes 5}$ is $\overrightarrow{OM}$
- the negative part of the amplitude $(-\frac{2}{\sqrt{2^5}}.|r\rangle)$ is modeled by $\overrightarrow{MN}$
- the $\overrightarrow{ON}$ models $|\psi\rangle = |p\rangle^{\otimes 5} - \frac{2}{\sqrt{2^5}}.|r\rangle$
- the vector $\overrightarrow{OH}$ models the projection of $|p\rangle^{\otimes 5}$ on $|r\rangle$
- the vector $\overrightarrow{OH'}$ models the projection of $|\psi\rangle$ on $|r\rangle$

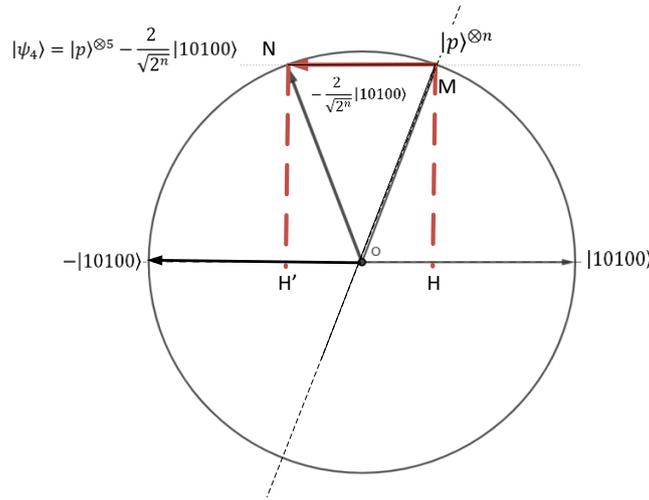

Figure 3. Visualization de $|\psi_4\rangle = |p\rangle^{\otimes 5} - \frac{2}{\sqrt{2^5}}.|10100\rangle$ at the end of Oracle

**Step 3.** Amplification

The amplitude amplification is interested in whether the state is in the subspace $E_1$ or not, and executes consecutive iterations starting from the initial state $|\psi_4\rangle$.

Considering any state $|\psi\rangle \in E$, amplitude amplifications refer to the reflexion operator as:
$$S_{|\Phi\rangle} = 2.|\Phi\rangle\langle\Phi| - Id^{\otimes n}$$
with $|\Phi\rangle \in E$
and one iteration of $S_{|\Phi\rangle}$ could be interpreted as one single Oracle execution.

$$S_{|\Phi\rangle}.|\psi\rangle = \sum_{x \in E}(-1)^{f(x)}.|x\rangle\langle x|$$

Consecutive iterations of amplitude amplification should gradually shift the distribution of amplitudes. We must apply the right number of iterations so that the resulting state is as close as possible of $|1\rangle^{\otimes n}$.

**Step 3.1**. Performing $H^{\otimes 5}$ permits to switch from $B(-|10100\rangle; |p\rangle^{\otimes 5})$ to basis $B\big(-|mpmpp\rangle; |0\rangle^{\otimes 5}\big)$ as stressed on figure 4.

$$|\psi_5\rangle = H^{\otimes 5}\left(|p\rangle^{\otimes 5} - \frac{2}{\sqrt{2^n}}.|10100\rangle\right)$$

$$|\psi_5\rangle = |0\rangle^{\otimes 5} - \frac{2}{\sqrt{2^n}}.H.|10100\rangle$$

Because $H.|0\rangle = |p\rangle$ and $H.|1\rangle = |m\rangle$ we have:

$$|\psi_5\rangle = |0\rangle^{\otimes 5} - \frac{2}{\sqrt{2^n}}.|mpmpp\rangle$$

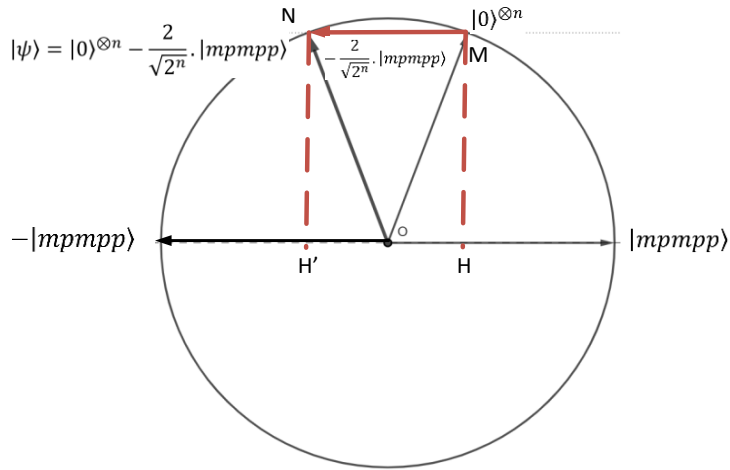

Figure 4. Visualization of $|\psi_5\rangle = |0\rangle^{\otimes 5} - \frac{2}{\sqrt{2^n}}.|mpmpp\rangle$

**Step 3.2.** Performing $X^{\otimes 5}$ permits to switch from $B\big(-|mpmpp\rangle; |0\rangle^{\otimes 5}\big)$ to $B(-|mpmpp\rangle; |1\rangle^{\otimes 5})$ as stressed on figure 5 where the plane is spanned by $|1\rangle^{\otimes n}$ and $-|mpmpp\rangle$.

$$|\psi_6\rangle = X^{\otimes 5}.\left(|0\rangle^{\otimes 5} - \frac{2}{\sqrt{2^n}}.|mpmpp\rangle\right)$$

$$|\psi_6\rangle = |1\rangle^{\otimes 5} - \frac{2}{\sqrt{2^n}}.|mpmpp\rangle$$

since $X.|p\rangle = |p\rangle$, $X.|m\rangle = -|m\rangle$ et $X.|0\rangle = |1\rangle$

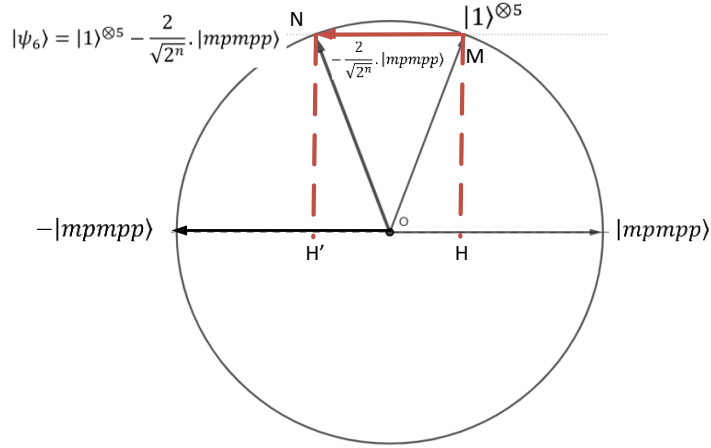

Figure 5. Visualization of $|\psi_6\rangle = |1\rangle^{\otimes 5} - \frac{2}{\sqrt{2^n}}.|mpmpp\rangle$

Since $X$ is an isometry (Raban and al., 2020) i.e. an inner-product preserving transformation that maps between two Hilbert spaces, the $X^{\otimes n}$ gate does not alter the relative position of $|\psi\rangle$ as regards both $X^{\otimes n}.|m\rangle^{\otimes n}$ and $X^{\otimes n}.|0\rangle^{\otimes n}$.

**Step 3.3.** Performing $CC \ldots CZ$

The geometric construction of $|\psi_6\rangle = |1\rangle^{\otimes 5} - \frac{2}{\sqrt{2^5}}.|mpmpp\rangle$ in figure 4 makes it clear that the $CCC \ldots CZ$ application should define a new state where $|\psi_7\rangle$ should be closest to $|1\rangle^{\otimes 5}$.

Performing $CCCCZ$ gives:

$$|\psi_7\rangle = CCCCZ.\left(|1\rangle^{\otimes 5} - \frac{2}{\sqrt{2^n}}.|mpmpp\rangle\right)$$

$$|\psi_7\rangle = -|1\rangle^{\otimes 5} - \frac{2}{\sqrt{2^5}}.CCCCZ(|mpmpp\rangle)$$

$$|\psi_7\rangle = -|1\rangle^{\otimes 5} - \frac{2}{\sqrt{2^5}}.\left(|mpmpp\rangle - \frac{2}{\sqrt{2^5}}.|11111\rangle\right)$$

$$|\psi_7\rangle = -|1\rangle^{\otimes 5} - \frac{2}{\sqrt{2^5}}.|mpmpp\rangle + \frac{4}{2^5}.|11111\rangle$$

$$|\psi_7\rangle = -|1\rangle^{\otimes 5} - \frac{2}{\sqrt{2^5}}.|mpmpp\rangle + \frac{1}{2^3}.|11111\rangle$$

$$|\psi_7\rangle = \left(\frac{1}{2^3} - 1\right).|1\rangle^{\otimes 5} - \frac{2}{\sqrt{2^5}}.|mpmpp\rangle$$

The quantum state is highlighted on figure 6.

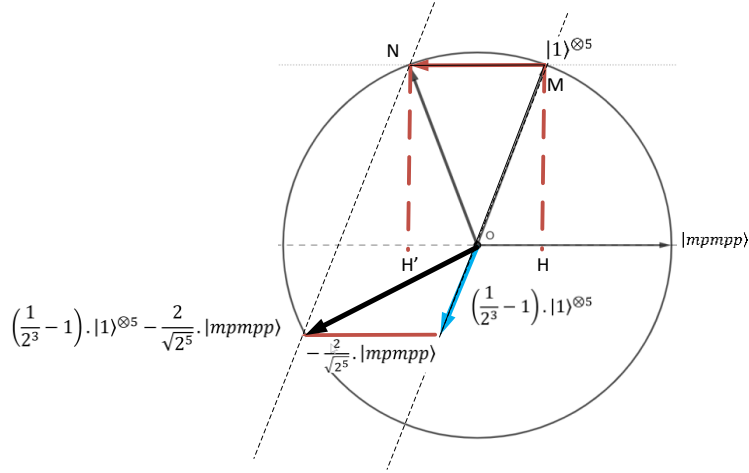

Figure 6. Visualization of $\left(\frac{1}{2^3} - 1\right).|1\rangle^{\otimes 5} - \frac{2}{\sqrt{2^5}}.|mpmpp\rangle$

**Step 3.4.** Performing $X^{\otimes 5}$

The $X^{\otimes 5}$ gate application at this point of the algorithm gives a $|\psi_8\rangle$ spanned by $|0\rangle^{\otimes 5}$ and $|pmpmm\rangle$ as stressed on figure 7.

$$|\psi_8\rangle = X^{\otimes 5}.\left[\left(\frac{1}{2^3} - 1\right).|1\rangle^{\otimes 5} - \frac{2}{\sqrt{2^5}}.|mpmpp\rangle\right]$$

$$|\psi_8\rangle = \left(\frac{1}{2^3} - 1\right).|0\rangle^{\otimes 5} - \frac{2}{\sqrt{2^5}}.|mpmpp\rangle$$

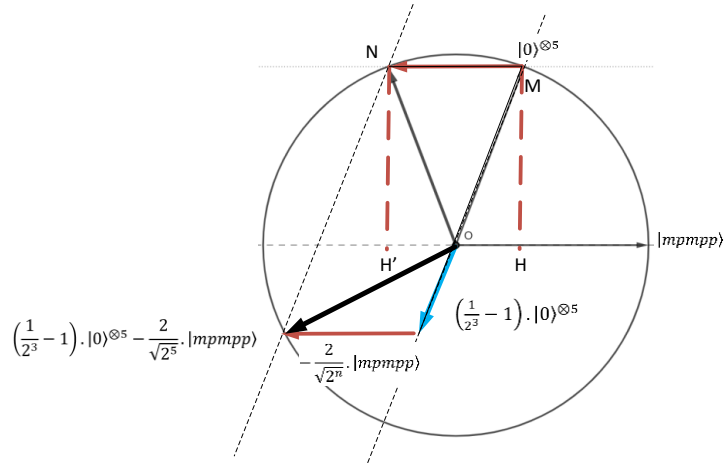

Figure 7. Visualization of $|\psi_8\rangle = \left(\frac{1}{2^3} - 1\right).|0\rangle^{\otimes 5} - \frac{2}{\sqrt{2^n}}.|mpmpp\rangle$

**Step 3.5.** Performing $H^{\otimes 5}$

The $H^{\otimes 5}$ gate application at this point of the algorithm makes a turn back to the initial bases $B(|10100\rangle; |p\rangle^{\otimes 5})$

$$|\psi_9\rangle = H^{\otimes 5} \cdot \left[\left(\frac{1}{2^3} - 1\right) \cdot |0\rangle^{\otimes 5} - \frac{2}{\sqrt{2^5}} \cdot |mpmpp\rangle\right]$$

$$|\psi_9\rangle = \left(\frac{1}{2^3} - 1\right) |p\rangle^{\otimes 5} - \frac{2}{\sqrt{2^5}} \cdot H \cdot |mpmpp\rangle$$

$$|\psi_9\rangle = \left(\frac{1}{2^3} - 1\right) |p\rangle^{\otimes 5} - \frac{2}{\sqrt{2^5}} \cdot |10100\rangle$$

The state $|\psi\rangle$ is defined by $\begin{pmatrix} \frac{1}{2^4} - 1 \\ -\frac{2}{\sqrt{2^5}} \end{pmatrix}$ in $B(|p\rangle^{\otimes 5}; |10100\rangle)$ as stressed on the figure 8, in the plane $P_{|p\rangle^{\otimes 5}; |10100\rangle}$.

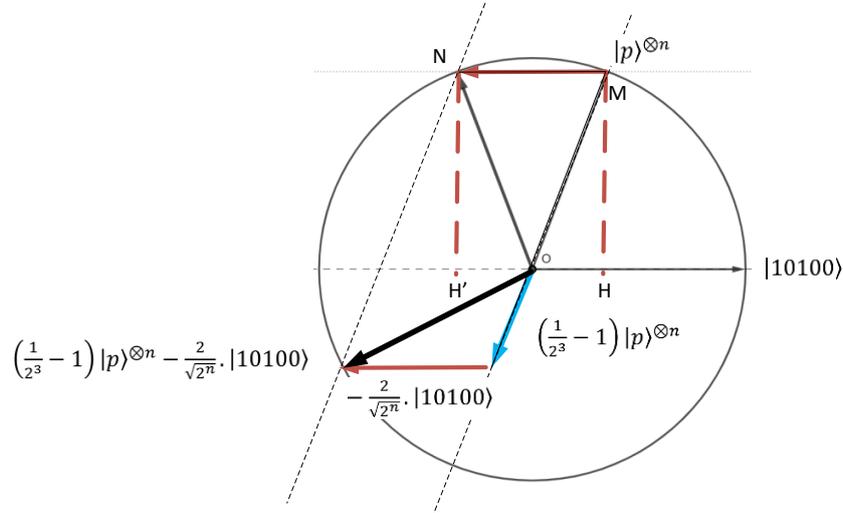

Figure 8. Visualization of $|\psi_9\rangle = \left(\frac{1}{2^3} - 1\right) |p\rangle^{\otimes 5} - \frac{2}{\sqrt{2^n}} \cdot |10100\rangle$

For a geometric point of view, we have drawn a parallelogram in the plane $B(|p\rangle^{\otimes 5}; |10100\rangle)$ as stressed on figure 9.

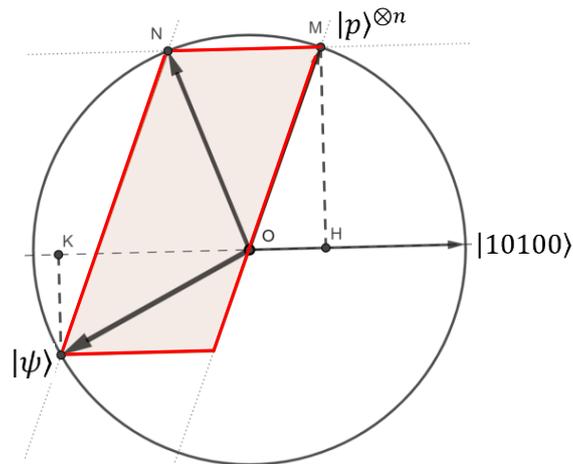

Figure 9. Visualization of $|\psi_9\rangle = \left(\frac{1}{2^3} - 1\right) |p\rangle^{\otimes 5} - \frac{2}{\sqrt{2^n}} \cdot |10100\rangle$

For $|\psi_9\rangle = \left(\frac{1}{2^3} - 1\right)|p\rangle^{\otimes 5} - \frac{2}{\sqrt{2^5}}.|10100\rangle$, then

$$\langle 10100|\psi_9\rangle = \left(\frac{1}{2^3} - 1\right)\langle 10100|p\rangle^{\otimes 5} - \frac{2}{\sqrt{2^5}}.\langle 10100|10100\rangle$$

$$\langle 10100|\psi_9\rangle = \left(\frac{1}{2^3} - 1\right).\frac{1}{\sqrt{2^5}} - \frac{2}{\sqrt{2^5}} = \left(\frac{1}{2^3} - 3\right).\frac{1}{\sqrt{2^5}} = \frac{23.\sqrt{2}}{64}$$

Hence the probability to find $|10100\rangle$ is
$$P\{|\psi\rangle = |10100\rangle\} = \frac{23^2 \times 2}{64^2} = \frac{529}{2048} \simeq 25.8\%$$
The initial probability $\left(\frac{1}{\sqrt{2^5}}\right)^2 = 3.1\%$ has been multiplied by 8.

*2.3 Grover geometric description for $|r\rangle^{\otimes n} = |1\rangle^{\otimes n}$.*

In the specific situation where $|r\rangle^{\otimes n} = |1\rangle^{\otimes n}$, the geometric representation of figure 10, shows that the state after the oracle is in the plane spanned by $|p\rangle^{\otimes n}$ and by $|1\rangle^{\otimes n}$.

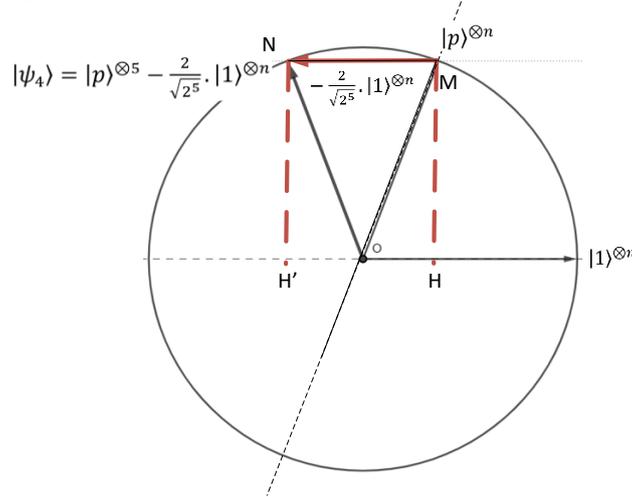

Figure 10. Visualization de $|\psi_4\rangle = |p\rangle^{\otimes 5} - \frac{2}{\sqrt{2^5}}.|1\rangle^{\otimes n}$ at the end of Oracle

**Step 3.** Amplification
Step 3.1. Performing $H^{\otimes n}$ permits to switch from $B(|1\rangle^{\otimes n}; |p\rangle^{\otimes n})$ to basis $B(|0\rangle^{\otimes n}; |m\rangle^{\otimes n})$ as stressed on figure 11.

$$|\psi_5\rangle = H^{\otimes n}\left(|p\rangle^{\otimes n} - \frac{2}{\sqrt{2^n}}.|1\rangle^{\otimes n}\right)$$

$$|\psi_5\rangle = |0\rangle^{\otimes n} - \frac{2}{\sqrt{2^n}}.|m\rangle^{\otimes n}$$

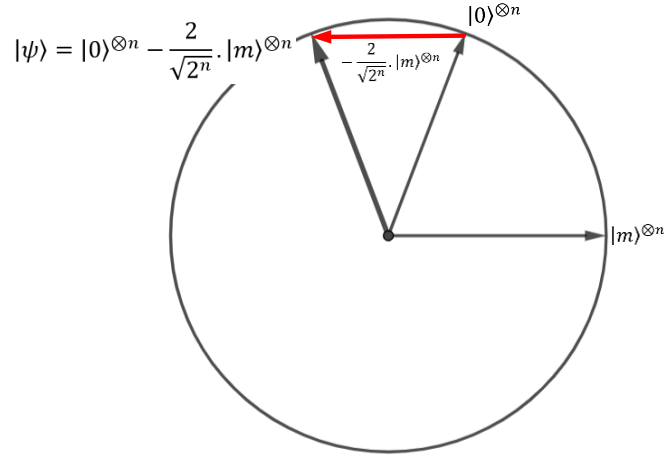

Figure 11. Visualization of $|\psi_5\rangle = |0\rangle^{\otimes n} - \frac{2}{\sqrt{2^n}}.|m\rangle^{\otimes n}$

Step 3.2. Performing $X^{\otimes n}$ permits to switch from $B(|m\rangle^{\otimes n}; |0\rangle^{\otimes n})$ to $B((-1)^n.|m\rangle^{\otimes n}; |1\rangle^{\otimes n})$ as stressed on figure 12 where the plane is spanned by $|1\rangle^{\otimes n}$ and $|m\rangle^{\otimes n}$.

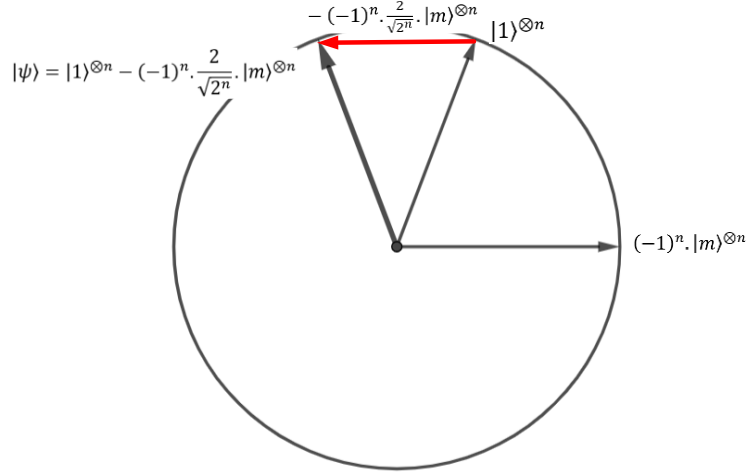

Figure 12. Visualization of $|\psi_6\rangle = |1\rangle^{\otimes n} - (-1)^n.\frac{2}{\sqrt{2^n}}.|m\rangle^{\otimes n}$

Note that for $n = 5$, we have $(-1)^n = (-1)^5 = -1$.

Step 3.3. Performing $CC \ldots CZ$

The geometric construction of $|\psi\rangle = |1\rangle^{\otimes n} - (-1)^n.\frac{2}{\sqrt{2^n}}.|m\rangle^{\otimes n}$ in figure 13 makes it clear that the $CCC \ldots CZ$ application should define a new state where $|\psi_6\rangle$ should be closest to $|1\rangle^{\otimes 5}$.

Performing $CC \ldots CZ$ gives (figure 12):

$$|\psi_7\rangle = CC \ldots CZ.\left(|1\rangle^{\otimes n} - (-1)^n.\frac{2}{\sqrt{2^n}}.|m\rangle^{\otimes n}\right)$$

$$|\psi_7\rangle = -|1\rangle^{\otimes n} - (-1)^n.\frac{2}{\sqrt{2^n}}.\left(|m\rangle^{\otimes n} - (-1)^n.\frac{2}{\sqrt{2^n}}.|1\rangle^{\otimes n}\right)$$

$$|\psi_7\rangle = -|1\rangle^{\otimes n} - (-1)^n \cdot \frac{2}{\sqrt{2^n}} \cdot |m\rangle^{\otimes n} + \frac{4}{2^n} \cdot |1\rangle^{\otimes n}$$

$$|\psi_7\rangle = \left(\frac{1}{2^{n-2}} - 1\right)|1\rangle^{\otimes n} - (-1)^n \cdot \frac{2}{\sqrt{2^n}} \cdot |m\rangle^{\otimes n}$$

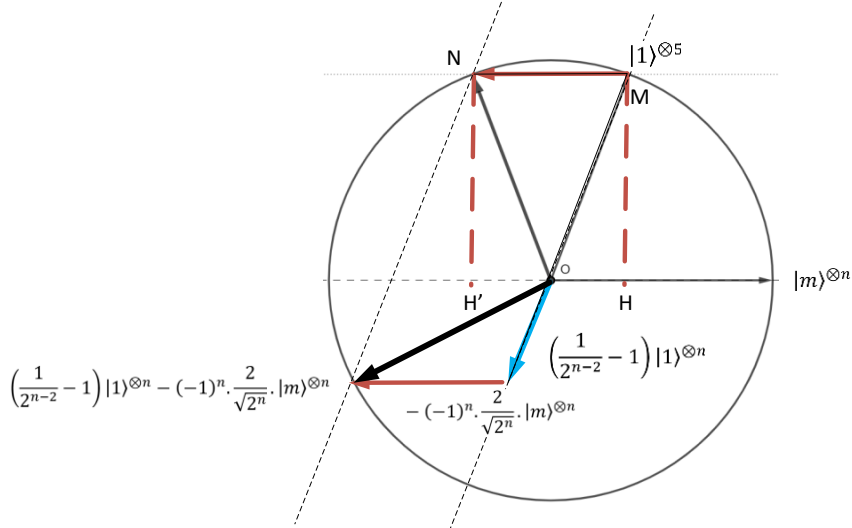

Figure 13. Visualization of $|\psi_7\rangle = \left(\frac{1}{2^{n-2}} - 1\right)|1\rangle^{\otimes n} - (-1)^n \cdot \frac{2}{\sqrt{2^n}} \cdot |m\rangle^{\otimes n}$

Step 3.4. Performing $X^{\otimes n}$
The $X^{\otimes n}$ gate application at this point of the algorithm gives a $|\psi\rangle$ spanned by $|m\rangle^{\otimes n}$ and $|0\rangle^{\otimes n}$ as stressed on figure 14.

$$|\psi_8\rangle = X^{\otimes n} \cdot \left[\left(\frac{1}{2^{n-2}} - 1\right)|1\rangle^{\otimes n} - (-1)^n \cdot \frac{2}{\sqrt{2^n}} \cdot |m\rangle^{\otimes n}\right]$$

$$|\psi_8\rangle = \left(\frac{1}{2^{n-2}} - 1\right)|0\rangle^{\otimes n} - \frac{2}{\sqrt{2^n}} \cdot |m\rangle^{\otimes n}$$

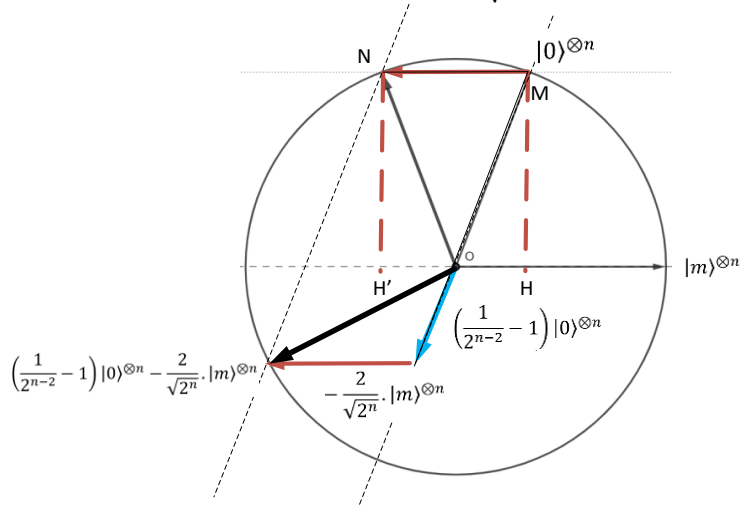

Figure 14. Visualization of $|\psi_8\rangle = \left(\frac{1}{2^{n-2}} - 1\right)|0\rangle^{\otimes n} - \frac{2}{\sqrt{2^n}} \cdot |m\rangle^{\otimes n}$

Step 3.5. Performing $H^{\otimes n}$

The $H^{\otimes n}$ gate application at this point of the algorithm gives a $|\psi_9\rangle$ spanned by $|1\rangle^{\otimes n}$ and $|p\rangle^{\otimes n}$ increasing the angle between $|\psi\rangle$ and $|1\rangle^{\otimes n}$ from $\mathrm{acos}\left(\frac{1}{\sqrt{2^n}}\right)$ to $\mathrm{acos}\left(\frac{2}{\sqrt{2^n}}\right)$.

$$|\psi_9\rangle = H^{\otimes n} \cdot \left[\left(\frac{1}{2^{n-2}} - 1\right)|0\rangle^{\otimes n} - \frac{2}{\sqrt{2^n}} \cdot |m\rangle^{\otimes n}\right]$$

$$|\psi_9\rangle = \left(\frac{1}{2^{n-2}} - 1\right)|p\rangle^{\otimes n} - \frac{2}{\sqrt{2^n}} \cdot |1\rangle^{\otimes n}$$

The state $|\psi_8\rangle$ is defined by $\begin{pmatrix}\frac{1}{2^{n-2}} - 1 \\ -\frac{2}{\sqrt{2^n}}\end{pmatrix}$ in $B(|1\rangle^{\otimes n}; |p\rangle^{\otimes n})$ as stressed on the figure 15, in the plane $P_{|1\rangle^{\otimes n}; |P\rangle^{\otimes n}}$.

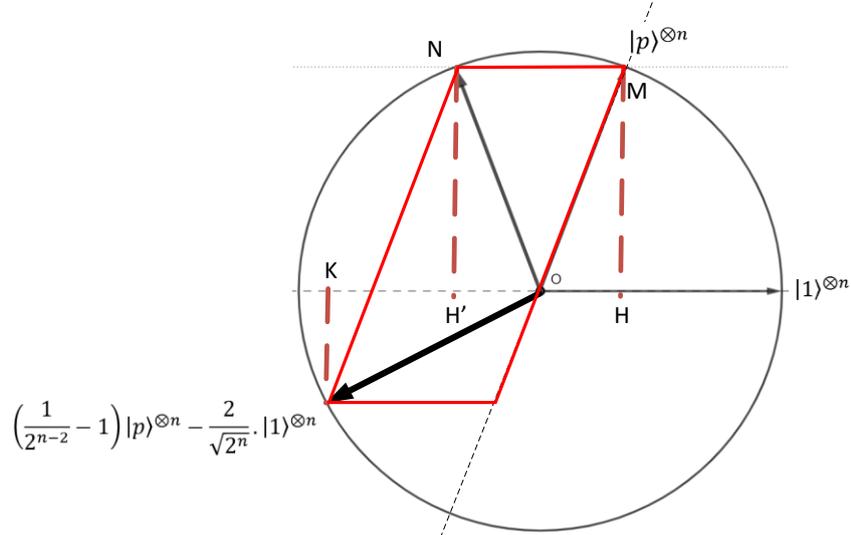

Figure 15. Visualization of $|\psi_9\rangle = \left(\frac{1}{2^{n-2}} - 1\right)|p\rangle^{\otimes n} - \frac{2}{\sqrt{2^n}} \cdot |1\rangle^{\otimes n}$

Note that $K$ on figure 15 is the orthogonal projection of $|\psi\rangle$ on $|1\rangle^{\otimes n}$ that is closer than the initial projection $H$.

**Concluding remarks**

The amplitude amplification has shifted the amplitude of $|p\rangle^{\otimes n}$ from $\frac{1}{\sqrt{2^n}}$ to $\left(\frac{1}{2^3} - 1\right)$

Because

$$|p\rangle^{\otimes n} = \frac{1}{\sqrt{2^n}} \cdot \left[\sum_{(e_1, e_2, \ldots, e_n) \in \{0;1\}^n} \frac{1}{\sqrt{2}} |e_1 \ldots e_n\rangle\right]$$

the probability of finding $|0\rangle^{\otimes n}$ has been updated from $\frac{1}{2^5}$ to $\left(\frac{1}{2^3} - 1\right)^2$.

$$|\psi\rangle = \left(\frac{1}{2^{n-2}} - 1\right)|p\rangle^{\otimes n} - \frac{2}{\sqrt{2^n}} \cdot |1\rangle^{\otimes n}$$

$$|\psi\rangle = \left(\frac{1}{2^{n-2}} - 1\right) \cdot \left(|p\rangle^{\otimes n} - \frac{1}{\sqrt{2^n}} \cdot |1\rangle^{\otimes n} + \frac{1}{\sqrt{2^n}} \cdot |1\rangle^{\otimes n}\right) - \frac{2}{\sqrt{2^n}} \cdot |1\rangle^{\otimes n}$$

$$|\psi\rangle = \left(\frac{1}{2^{n-2}} - 1\right) \cdot \left(|p\rangle^{\otimes n} - \frac{1}{\sqrt{2^n}} \cdot |1\rangle^{\otimes n}\right) + \left(\frac{1}{2^{n-2}} - 1\right) \cdot \frac{1}{\sqrt{2^n}} \cdot |1\rangle^{\otimes n} - \frac{2}{\sqrt{2^n}} \cdot |1\rangle^{\otimes n}$$

$$|\psi\rangle = \left(\frac{1}{2^{n-2}} - 1\right) \cdot \left(|p\rangle^{\otimes n} - \frac{1}{\sqrt{2^n}} \cdot |1\rangle^{\otimes n}\right) + \left[\left(\frac{1}{2^{n-2}} - 1\right) \cdot \frac{1}{\sqrt{2^n}} \cdot - \cdot \frac{2}{\sqrt{2^n}}\right] |1\rangle^{\otimes n}$$

The state $|\psi\rangle$ is spanned by $\left(|p\rangle^{\otimes n} - \frac{1}{\sqrt{2^n}} \cdot |1\rangle^{\otimes n}\right)$ and by $|1\rangle^{\otimes n}$ and the basis is now $B\left(\left(|p\rangle^{\otimes n} - \frac{1}{\sqrt{2^n}} \cdot |1\rangle^{\otimes n}\right); |1\rangle^{\otimes n}\right)$.

The measurement outcome of $|\psi\rangle$ is:

$$P(|\psi\rangle = |1\rangle) = \left[\left(\frac{1}{2^{n-2}} - 1\right) \cdot \frac{1}{\sqrt{2^n}} \cdot - \cdot \frac{2}{\sqrt{2^n}}\right]^2$$

and $P(|\psi\rangle = |1\rangle)$ characterizes how the state is collapsed on $|1\rangle$ by a measurement. The quantum state throughout its evolution gives for one $|\psi\rangle \in E_1$ a probability $P(|\psi\rangle = |1\rangle)$ most probable throughout its evolution from $\frac{1}{2^n}$ to $\left[\left(\frac{1}{2^{n-2}} - 1\right) \cdot \frac{1}{\sqrt{2^n}} \cdot - \cdot \frac{2}{\sqrt{2^n}}\right]$ and gives for one $|\psi\rangle \in E_2$ a probability $P(|\psi\rangle = |1\rangle)$ less probable.

In the very specific situation where $Card(E_1) = 1$ and $n = 5$ we have $P(|\psi\rangle = |1\rangle) \simeq 25\%$

## 2.3. Conclusion for one $|r\rangle^k = \otimes_{j=1,n} |r_j\rangle$

To conclude, the algorithm is based on 3 main steps: the first one is the initialization, the second one the Oracle and the last one the amplification.

**Step 1. Initialization**
  **Step 1.0.** $|\psi_0\rangle = |0\rangle^{\otimes n} \otimes |1\rangle$
  **Step 1.1.** Application of $H^{\otimes n+1}$

$$|\psi_1\rangle = |p\rangle^{\otimes n} \otimes |m\rangle$$

**Step 2. Oracle**
The Oracle consists in performing controlled $X$ to change the phase of all $|r\rangle^k \in E_1$ by $\pi$ to defined the quantum state:

$$|\psi\rangle = |p\rangle^{\otimes n} - \frac{2}{\sqrt{2^n}} \cdot \sum_{|r\rangle^k \in E_1} |r\rangle^k$$

$$|\psi\rangle = |p\rangle^{\otimes n} - \frac{2}{\sqrt{2^n}} \cdot \sum_{|r\rangle^k \in E_1} |r\rangle^k$$

The state $|\psi\rangle$ is in the plane spanned by $|p\rangle^{\otimes n}$ and by all $|r\rangle^k = \otimes_{j=1,n} |r_j\rangle$ with $|r\rangle^k \in E_1$. The basis is $B(|r\rangle^k; |p\rangle^{\otimes n})$.

Steps of the Oracle

**Step 2.1.** The **first step** of the Oracle consists in switching the base $B(|r\rangle^k ; |p\rangle^{\otimes n})$ into $B(|1\rangle^{\otimes n}; |p\rangle^{\otimes n})$ that is the only base where it possible to achieved the $CX$.

**Step 2.2.** The **second step** consist in application of $CX$.

The $CX$ changes the component on $|1\rangle^{\otimes n}$ into its opposite. By consequence the basis is switched to:
$$|\psi\rangle = |p\rangle^{\otimes n} - \frac{2}{\sqrt{2^n}} \cdot |1\rangle^{\otimes n}$$

**Step 2.3.** The **third step** of the Oracle consist in turning back into $B(|r\rangle^k ; |p\rangle^{\otimes n})$

**Step 3. Amplification**

**Step 3.1**. Performing $H^{\otimes n}$ permits to switch from $B(|r\rangle^k ; |p\rangle^{\otimes n})$ to basis $B([H^{\otimes n} \cdot |r\rangle^k]; |0\rangle^{\otimes n})$

**Step 3.2.** Performing $X^{\otimes n}$ permits to switch from $B([H^{\otimes n} \cdot |r\rangle^k]; |0\rangle^{\otimes n})$ to $B([X^{\otimes n} \cdot H^{\otimes n} \cdot |r\rangle^k]; |1\rangle^{\otimes n})$

**Step 3.3**. Performing $CC \ldots CZ$ to change the component on $|1\rangle^{\otimes n}$ into its opposite.

Let us denote $c^k$ the component of $X^{\otimes n} \cdot H^{\otimes n} \cdot |r\rangle^k$ on $|1\rangle^{\otimes n}$ and we have:
$$CC \ldots CZ \cdot (X^{\otimes n} \cdot H^{\otimes n} \cdot |r\rangle^k) = X^{\otimes n} \cdot H^{\otimes n} \cdot |r\rangle^k - 2 \cdot c^k \cdot |1\rangle^{\otimes n}$$
By consequence the basis is switched to:
$$B([X^{\otimes n} \cdot H^{\otimes n} \cdot |r\rangle^k - 2 \cdot c^k \cdot |1\rangle^{\otimes n}] ; |1\rangle^{\otimes n})$$

**Step 3.4.** Performing $X^{\otimes n}$ permits to switch back to $B([H^{\otimes n} \cdot |r\rangle^k - 2 \cdot c^k \cdot |0\rangle^{\otimes n}] ; |0\rangle^{\otimes n})$

**Step 3.5.** Performing $H^{\otimes n}$ permits to switch back to $B([|r\rangle^k - 2 \cdot c^k \cdot |p\rangle^{\otimes n}] ; |p\rangle^{\otimes n})$

For a geometric point of view, one can draw a parallelogram in the plane $B(|r\rangle; |p\rangle^{\otimes n})$ as stressed on figure 15. The "action of Oracle" consists is subtracting to $|p\rangle^{\otimes n}$ a part of $|r\rangle$ that is modeled by a left red vector in figure 16. The "Action of amplification" is a subtraction of a part of $|p\rangle^{\otimes n}$.

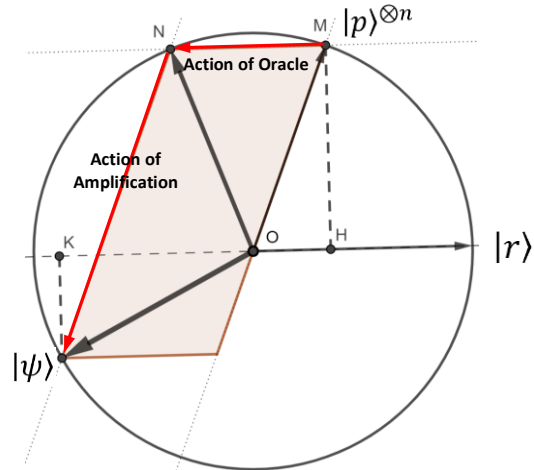

Figure 16. Visualization of $|\psi\rangle$ after one amplification

The second iteration starts with the Oracle followed by one amplification as stressed on figure 16. Because $OL > OK$ the probability $P(|\psi\rangle = |r\rangle)$ has been improved by the second amplification (figure 17).

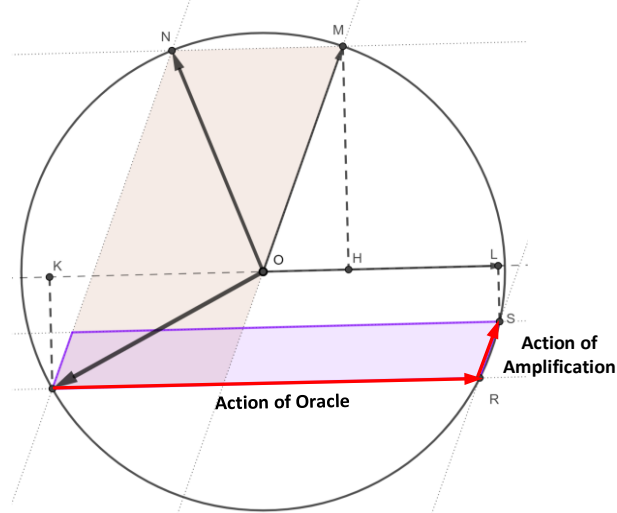

Figure 17. Visualization of $|\psi\rangle$ after the second amplification

The third is unprofitable because the $P(|\psi\rangle = |r\rangle)$ decreases as stressed on figure 18 where $OJ < OL$

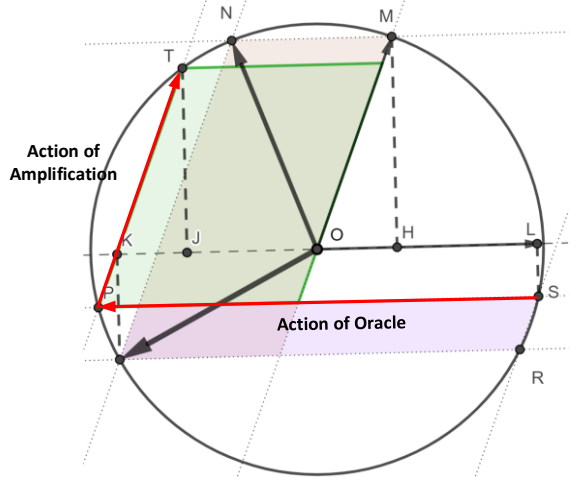

Figure 18. Visualization of $|\psi\rangle$ after the third amplification

The amplification procedure increases the probability of $|r\rangle^k$ according to the iterative process described above.

Let us note $\theta / \acos\left(\frac{1}{\sqrt{2^n}}\right) = \theta$ i.e. the probability of each state is $\left(\frac{1}{\sqrt{2^n}}\right)^2 = \frac{1}{2^n}$. Each amplification increases of $2.\theta \pmod{2\pi}$ the marked current quantum states. Hence the phase of solution is successively: $3.\theta \pmod{2\pi}$, $5.\theta \pmod{2\pi}$, $7.\theta \pmod{2\pi}$,…

It is important to analyze not only $P(|\psi\rangle = |r\rangle)$ but the difference between $P(|\psi\rangle = |r\rangle)$ (that is the probability to find $|r\rangle \in E_1$) and $P(|\psi\rangle = |s\rangle)$ with $|s\rangle \in E_0$. For example, with $n = 5$ and $\#E_1 = 1$, after one iteration, $P(|\psi\rangle = |r\rangle) = 25.8\%$ and the probability of each $|s\rangle \in E_0$ is ten times lower (about 2.5%) giving a very significant ratio between probability of states in $E_1$ and states in $E_2$ about 10 times.

Let us note that at iteration 1, $P(|\psi\rangle = |r\rangle)$ has been increased from 3.1% to 25.8% as stressed on table 1. These values meet the numerical experiments achieved with Qiskit and available in the Appendix.

Table 1. $n = 5$ and $\#E_1 = 1$

| Iteration | $\theta$ | $\cos(\theta)$ | $P\{|\psi\rangle = |s\rangle\ /\ |s\rangle \in E_0\}$ | $P\{|\psi\rangle \in E_1\}$ |
|---|---|---|---|---|
| 0 | $\theta = 1.393$ | 0.177 | 0.031 | **0.031** |
| 1 | $3.\theta = 4.179$ | -0.508 | **0.024** | **0.258** |
| 2 | $5.\theta = 0.682$ | 0.776 | 0.013 | 0.602 |
| 3 | $7.\theta = 3.468$ | -0.947 | 0.003 | 0.897 |
| 4 | $9.\theta = 6.255$ | 1.000 | 0.000 | 0.999 |
| 5 | $11.\theta = 2.758$ | -0.927 | 0.005 | 0.860 |

## 3. A Grover circuit with 3 qubits

In this section we give a circuit version based on 3 qubits only, and we analyzed its evolution using simple geometric consideration. At each step we point out how the basis changes trying to give an intuitive representation of the quantum states.

This figure 19 describes a classical circuit based on Grover amplification where the Oracle has marked the $|10\rangle$ state by:

$$(X \otimes Id^{\otimes 2}).CC \dots CNOT_{1,2;3.}(X \otimes Id^{\otimes 2})$$

where $CCNOT_{1,2;3.}$ is the conditional $X - gate$ from 1 and 2 to 3.

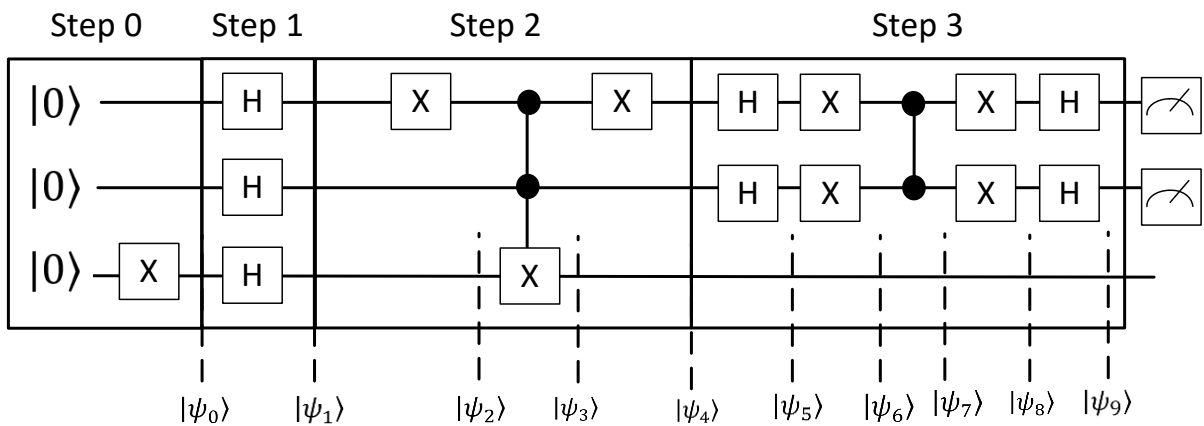

Figure 19. One basic circuit with the Grover's amplification procedure

**Step 1. Initialization**

**Step 1.0.**

The initial quantum state is $|\psi_0\rangle = |001\rangle$

where the qubit number 3 is $|1\rangle$.

**Step 1.1.** Because $H.|0\rangle = |p\rangle$ and $H.|1\rangle = |m\rangle$, then the current state is spanned by $|pp\rangle$ and $|m\rangle$ :

$$H^{\otimes 3}.|001\rangle = |pp\rangle \otimes |m\rangle$$

The operator $H^{\otimes 3}$ applied to $|\psi_0\rangle$ gives an equal superposition of the states in the computational base resulting in outcome $i$ with an amplitude about $\frac{1}{\sqrt{2^3}}$ for half of them and about $-\frac{1}{\sqrt{2^3}}$ for the last ones.

$$|\psi_1\rangle = H^{\otimes 3}.|001\rangle$$

$$|\psi_1\rangle = \frac{1}{2\sqrt{2}}(|000\rangle - |001\rangle + |010\rangle - |011\rangle + |100\rangle - |101\rangle + |110\rangle - |111\rangle)$$

**Step 2. Oracle definition**

Step 2.1. Application of $(X \otimes Id^{\otimes 2})$

$$|\psi_2\rangle = (X \otimes Id^{\otimes 2}).(|pp\rangle \otimes |m\rangle) = |ppm\rangle$$

defining the state:

$$|\psi_2\rangle = \frac{1}{2\sqrt{2}}(|000\rangle - |001\rangle + |010\rangle - |011\rangle + |100\rangle - |101\rangle + |110\rangle - |111\rangle)$$

Step 2.2. Application of $CCX$

Application of $CCX_{1,2;3}$ consists in application of $X$ on qubit number 3 when both qubits 1 and 2 are $|1\rangle$ meaning that $CCX_{1,2;3}.|110\rangle = |111\rangle$ and $CCX_{1,2;3}.-|111\rangle = -|110\rangle$.

So

$$|\psi_3\rangle = CCX_{1,2;3}.|ppm\rangle$$

i.e.

$$|\psi_3\rangle = \frac{1}{2\sqrt{2}}(|000\rangle - |001\rangle + |010\rangle - |011\rangle + |100\rangle - |101\rangle + |111\rangle - |110\rangle)$$

The operator in practice has changed the amplitude of both $|110\rangle$ and $|111\rangle$. The $|110\rangle$ amplitude has been changed from $\frac{1}{2\sqrt{2}}$ to $-\frac{1}{2\sqrt{2}}$ and the $|111\rangle$ amplitude has been changed from $-\frac{1}{2\sqrt{2}}$ to $\frac{1}{2\sqrt{2}}$.

Because

$$-\frac{2}{2\sqrt{2}}|110\rangle + \frac{2}{2\sqrt{2}}|111\rangle = -|11\rangle \otimes \left(\frac{1}{\sqrt{2}}|0\rangle + \frac{1}{\sqrt{2}}|1\rangle\right) = -|11\rangle \otimes |m\rangle$$

we have

$$|\psi_3\rangle = CCX_{1,2;3}.|ppm\rangle$$
$$|\psi_3\rangle = |ppm\rangle - |11m\rangle$$
$$|\psi_3\rangle = (|pp\rangle - |11\rangle) \otimes |m\rangle$$

Step 2.3. Application of $X \otimes Id \otimes Id$

Knowing that $X.|p\rangle = |p\rangle$, $X.|1\rangle = |0\rangle$ and $X.|0\rangle = |1\rangle$ we have:

$$|\psi_4\rangle = (X \otimes Id \otimes Id).(|ppm\rangle - |11m\rangle)$$
$$|\psi_4\rangle = |ppm\rangle - |01m\rangle$$
$$|\psi_4\rangle = (|pp\rangle - |01\rangle) \otimes |m\rangle$$

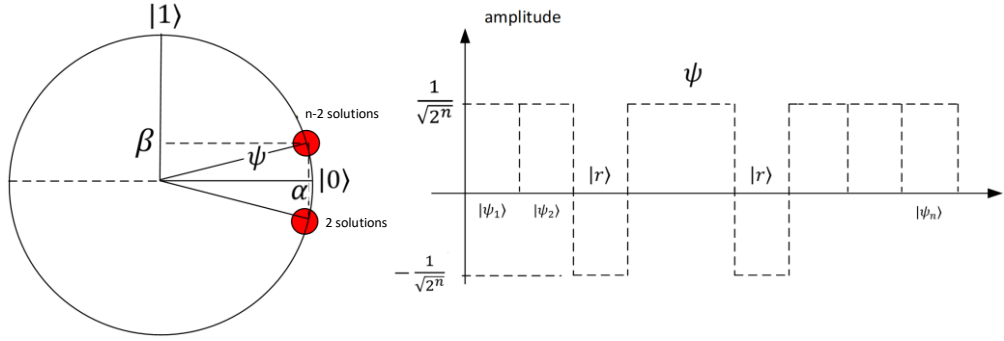

Figure 20. $|\psi_4\rangle$ in the computational base (base of $Z^{\otimes 2} - gate$)

Assigned a similar amplitude to all states in $B(|0\rangle; |1\rangle)$. In the plane spanned by $|p\rangle^{\otimes 2}$ and $|m\rangle^{\otimes 2}$, the state $|010\rangle$ and $|011\rangle$ are the marked elements with $-1$. In the plane spanned by $|0\rangle^{\otimes 3}$ and $|1\rangle^{\otimes 3}$ the probabilities of all elements are identical and 2 of them have now negative amplitude. The principle is illustrated on figure 20 where a part of the amplitude has now negative value.

Without loss of generality the qubit $|m\rangle$ could be vanished since, during the next steps of the algorithm, only $Id - gate$ are performed on this qubit. So it is possible to consider the following $|\psi_4\rangle$ definition only:

$$|\psi_4\rangle = (|pp\rangle - |01\rangle)$$

that models $|\psi_4\rangle = \frac{1}{2}(|00\rangle + |01\rangle - |10\rangle + |11\rangle)$ and to provide a readable geometric representation on figure 19.

**Step 3. Amplification**

Step 3.1. Application of $H^{\otimes 2} \otimes Id$

$$|\psi_5\rangle = (H^{\otimes 2} \otimes Id).(|ppm\rangle - |01m\rangle) = |00m\rangle - |pmm\rangle = (|00\rangle - |pm\rangle) \otimes |m\rangle$$

$|\psi_4\rangle$ is now in the plane spanned by $|00\rangle$ and $|pm\rangle$ (figure 20) since $H^{\otimes 2}$ has permit to switch from $B(|pp\rangle; |01\rangle)$ to $B(|00\rangle; -|pm\rangle)$ (figure 21).

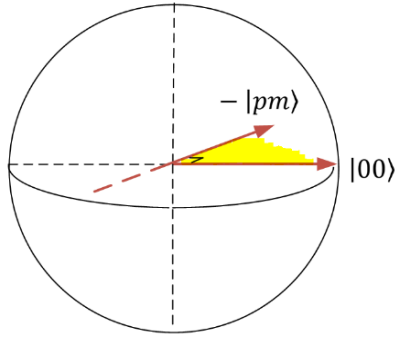

Figure 21. Basis $B(|00\rangle; -|pm\rangle)$

Step 3.2. Application of $X^{\otimes 2} \otimes Id$

Since $X.|p\rangle = |p\rangle$ and $X.|m\rangle = -|m\rangle$, we have:

$$|\psi_6\rangle = (X^{\otimes 2} \otimes Id).[(|00\rangle - |pm\rangle) \otimes |m\rangle]$$

$$|\psi_6\rangle = (|11\rangle + |pm\rangle) \otimes |m\rangle$$

$|\psi_6\rangle$ is in the plane spanned by $|11\rangle$ and $|pm\rangle$ since $X^{\otimes 2}$ switches from $B(|00\rangle; -|pm\rangle)$ to $B(|11\rangle; |pm\rangle)$ (figure 22).

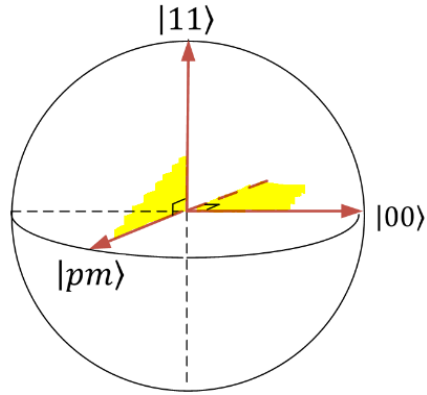

Figure 22. Basis $B(|11\rangle; |pm\rangle)$

Step 3.3. Application de $CZ_{1;2} \otimes Id$

$$|\psi_7\rangle = (CZ_{1;2} \otimes Id).(|11\rangle + |pm\rangle) \otimes |m\rangle = CZ_{1;2}.(|11\rangle + |pm\rangle) \otimes |m\rangle$$

We have

$$CZ_{1;2}.\left[\frac{1}{2}(|00\rangle + |01\rangle + |10\rangle - |11\rangle)\right] = \frac{1}{2}.(|00\rangle + |01\rangle + |10\rangle + |11\rangle)$$

and by consequence: $CZ_{1;2}.|pm\rangle = |pm\rangle + |11\rangle$

So

$$|\psi_7\rangle = (-|11\rangle + |pm\rangle + |11\rangle) \otimes |m\rangle$$

$$|\psi_7\rangle = |pm\rangle \otimes |m\rangle$$

So $CZ_{1;2} - gate$ defines a quantum state that is fully composed of $|pm\rangle$ (figure 23).

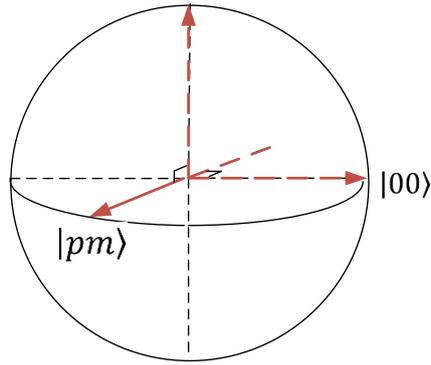

Figure 23. Visualization of $|\psi_5\rangle$

Etape 3.4. Application de $X^{\otimes 2} \otimes Id$

$$|\psi_8\rangle = (X^{\otimes 2} \otimes Id).|\psi_5\rangle$$
$$|\psi_8\rangle = (X^{\otimes 2} \otimes Id).(|pm\rangle \otimes |m\rangle)$$
$$|\psi_8\rangle = -|pm\rangle \otimes |m\rangle$$

The quantum state is represented on figure 24.

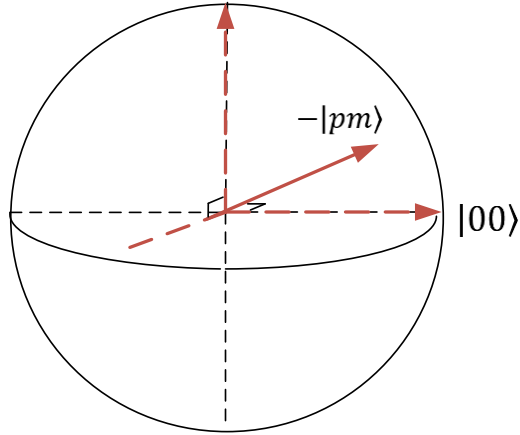

Figure 24. Visualization of $|\psi_6\rangle$

Etape 3.5. Application de $H^{\otimes 2} \otimes Id$

The amplitude algorithm executes $H^{\otimes 2} \otimes Id$ allowing at the end a measurement in the computational basis (figure 25).

$$|\psi_9\rangle = (H^{\otimes 2} \otimes Id).|\psi_6\rangle$$
$$|\psi_9\rangle = (H^{\otimes 2} \otimes Id).(-|pm\rangle \otimes |m\rangle)$$
$$|\psi_9\rangle = -|01\rangle \otimes |m\rangle$$

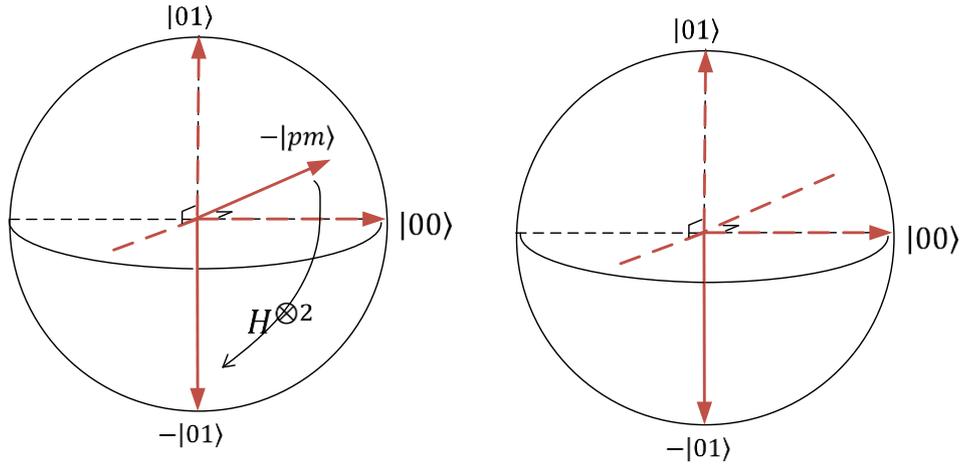

Figure 25. Visualization of $|\psi_7\rangle$

In the computational $B(|00\rangle; ...; |11\rangle)$, the amplitude of $|01\rangle$ is 1 meaning that applying a measurement on the computational base outcome is $|01\rangle$ with a probability of 100%. In summary, all computations lie on 8 transitional states that are sum-up in the table 1.

Table 3. All successive quantum states during Grover's based circuit

| | Quantum State |
|---|---|
| Step 0 | $|\psi_0\rangle = |001\rangle$ |
| Step 1. Application of $H^{\otimes 3}$ | $|\psi_1\rangle = |pp\rangle \otimes |m\rangle$ |
| **Oracle** | |
| Step 2.1. Application of $X \otimes Id \otimes Id$ | $|\psi_2\rangle = |ppm\rangle$ |
| Step 2.2. Application of $CCX(q_1, q_2; q_3)$ | $|\psi_3\rangle = |ppm\rangle - |11m\rangle$ |
| Step 2.2. Application of $X \otimes Id \otimes Id$ | $|\psi_4\rangle = |ppm\rangle - |01m\rangle = (|pp\rangle - |01\rangle) \otimes |m\rangle$ |
| **Amplification** | |
| Step 3.1. Application of $H^{\otimes 2} \otimes Id$ | $|\psi_5\rangle = (|00\rangle - |pm\rangle) \otimes |m\rangle$ |
| Step 3.2. Application of $X^{\otimes 2} \otimes Id$ | $|\psi_6\rangle = (|11\rangle + |pm\rangle) \otimes |m\rangle$ |
| Step 3.3. Application of $CZ(q_1; q_2) \otimes Id$ | $|\psi_7\rangle = [-|11\rangle + (|pm\rangle + |11\rangle)] \otimes |m\rangle$ <br> $|\psi_7\rangle = (-|11\rangle + |11\rangle + |pm\rangle) \otimes |m\rangle$ <br> $|\psi_7\rangle = |pm\rangle \otimes |m\rangle$ |
| Step 3.4. Application of $X^{\otimes 2} \otimes Id$ | $|\psi_8\rangle = -|pm\rangle \otimes |m\rangle$ |
| Step 3.5. Application of $H^{\otimes 2} \otimes Id$ | $|\psi_9\rangle = -|01\rangle \otimes |m\rangle$ |

## 4. A Grover circuit with two amplifications

Let us consider a Grover-based circuit with 3 qubits using a $CZ - gate$ for the Oracle and let us assume that $E_1 = \{|001\rangle\}$. The last qubit is the auxiliary qubits (referred to as ancilla qubit for example in (Grover, 2000)) required for the $CZ - gate$ and to obtain more readable state we do not model this qubit during the computation of successive quantum states.

The figure 26 gives the circuit that takes advantages of $CZ - gate$ on both Oracle part and amplification which is a alternative to the classical Grover circuit.

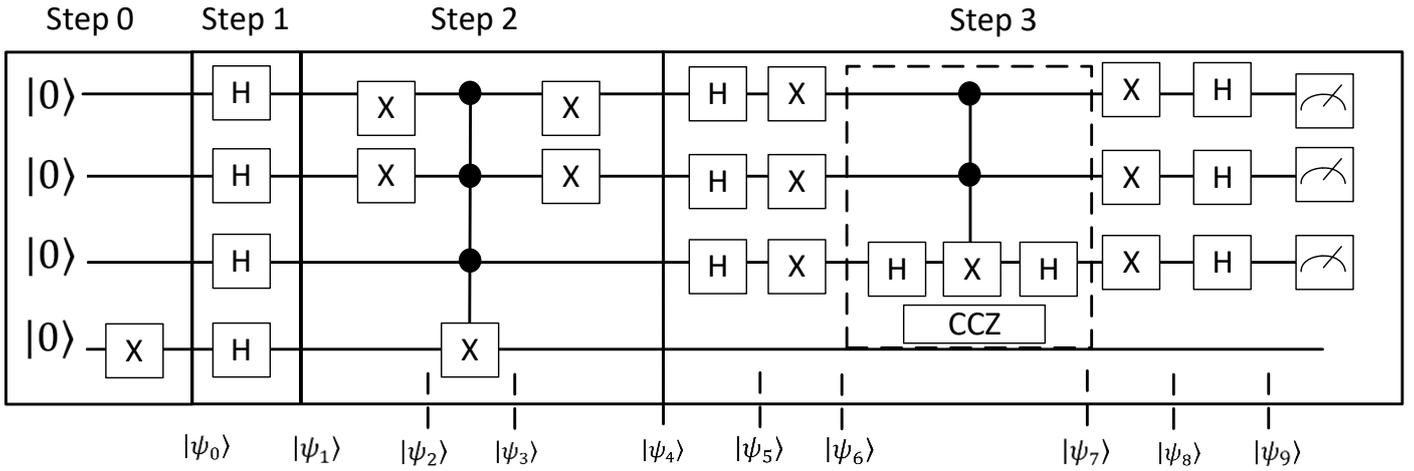

Figure 26. Example of one Grover-based circuit composed of 3-qubits and one amplification

$H^{\otimes 3} - gate$ is performed at step 3.5.

$$|\psi_9\rangle = H^{\otimes 3}.\left(-\frac{1}{\sqrt{2}}.|ppm\rangle - \frac{1}{2}.|000\rangle\right)$$

$$|\psi_9\rangle = -\frac{1}{\sqrt{2}}.|001\rangle - \frac{1}{2}.|ppp\rangle$$

Table 4. Successive quantum states during 3 qubits Grover's based circuit

|  | Quantum State |
|---|---|
| Step 0. | $|\psi_0\rangle = |000\rangle$ |
| Step 1. Application of $H^{\otimes 3}$ | $|\psi_1\rangle = |ppp\rangle$ |
| **Oracle** | |
| Step 2.1. Application of $X^{\otimes 2} \otimes Id$ | $|\psi_2\rangle = |ppp\rangle$ |
| Step 2.2. Application of $CZ(q_1, q_2; q_3)$ | $|\psi_3\rangle = |ppp\rangle - \frac{1}{\sqrt{2}}.|111\rangle$ |
| Step 2.3. Application of $X^{\otimes 2} \otimes Id$ | $|\psi_4\rangle = |ppp\rangle - \frac{1}{\sqrt{2}}.|001\rangle$ |
| **Amplification** | |
| Step 3.1. Application of $H^{\otimes 3}$ | $|\psi_5\rangle = |000\rangle - \frac{1}{\sqrt{2}}.|ppm\rangle$ |
| Step 3.2. Application of $X^{\otimes 3}$ | $|\psi_6\rangle = |111\rangle + \frac{1}{\sqrt{2}}.|ppm\rangle$ |
| Step 3.3. Application of $CZ(q_1, q_2; q_3)$ | $|\psi_7\rangle = -|111\rangle + \frac{1}{\sqrt{2}}.\left(|ppm\rangle + \frac{1}{\sqrt{2}}.|111\rangle\right)$ $= \frac{1}{\sqrt{2}}.|ppm\rangle - \frac{1}{2}.|111\rangle$ |
| Step 3.4. Application of $X^{\otimes 3}$ | $|\psi_8\rangle = -\frac{1}{\sqrt{2}}.|ppm\rangle - \frac{1}{2}.|000\rangle$ |
| Step 3.5. Application of $H^{\otimes 3}$ | $|\psi_9\rangle = -\frac{1}{\sqrt{2}}.|001\rangle - \frac{1}{2}.|ppp\rangle$ |

But $\langle 001|ppp\rangle = \frac{1}{2.\sqrt{2}}$ hence $\langle 001|\psi_9\rangle = -\frac{1}{2}\langle 001|ppp\rangle - \frac{1}{\sqrt{2}}.\langle 001|001\rangle = -\frac{1}{4.\sqrt{2}} - \frac{1}{\sqrt{2}}$

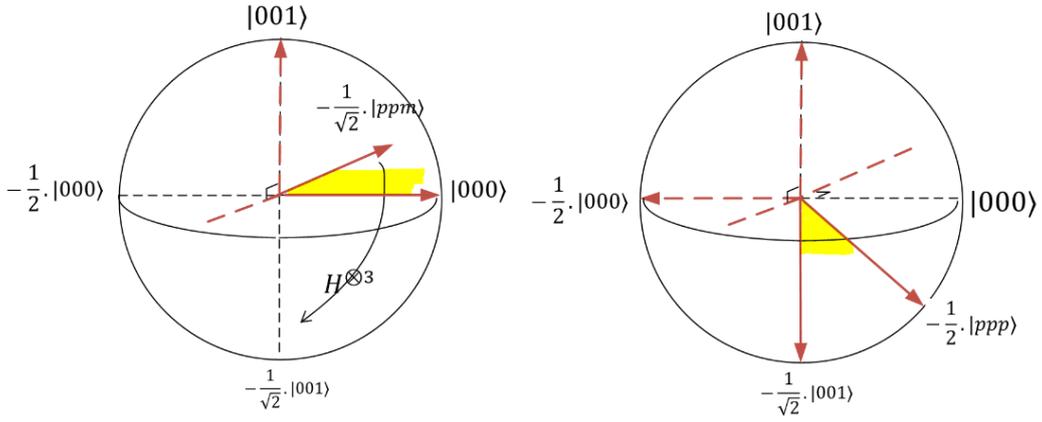

Figure 26. Switch to the computational basis using $H^{\otimes 3}$

Hence

$$P\{|\psi_9\rangle = |001\rangle\} = \left[-\frac{1}{\sqrt{2}} - \frac{1}{2} \times \frac{1}{2\sqrt{2}}\right]^2 = \left[\frac{1}{\sqrt{2}} + \frac{1}{4\sqrt{2}}\right]^2 = \left[\frac{5}{4\sqrt{2}}\right]^2 = \frac{25}{32} \simeq 78.1\%$$

meaning that the probability to obtain $|001\rangle$ is about 78%.

The numerical experiments achieved on the Qiskit composer (IBM) meet the theoretical considerations.

The second amplification is described on the table 5 and starts with the state $|\psi_9\rangle = -\frac{1}{\sqrt{2}}.|001\rangle - \frac{1}{2}.|ppp\rangle$ and provides at the end the quantum state: $|\psi_8\rangle = \frac{3}{2\sqrt{2}}.|001\rangle - \frac{1}{4}.|ppp\rangle$. The measurements should yield a probability $P(|\psi_8\rangle = |001\rangle)$ that can be easily evaluated by $P(|\psi_8\rangle = |001\rangle) = \left[+\frac{3}{2\sqrt{2}} - \frac{1}{4} \times \frac{1}{2\sqrt{2}}\right]^2$ and

$$P(|\psi_8\rangle = |001\rangle) = \left[\frac{3}{2\sqrt{2}} - \frac{1}{8\sqrt{2}}\right]^2 = \left[\frac{11}{8\sqrt{2}}\right]^2 \simeq 97\%$$

Table 5. Second amplification

|  | **Quantum State** |
|---|---|
|  | $|\psi_1\rangle = -\frac{1}{\sqrt{2}}.|001\rangle - \frac{1}{2}.|ppp\rangle$ |
|  | **Oracle** |
| Step 2.1. Application of $X^{\otimes 2} \otimes Id$ | $|\psi_2\rangle = -\frac{1}{\sqrt{2}}.|111\rangle - \frac{1}{2}.|ppp\rangle$ |
| Step 2.2. Application of $CZ(q_1, q_2; q_3)$ | $|\psi_3\rangle = \frac{1}{\sqrt{2}}.|111\rangle - \frac{1}{2}.\left(|ppp\rangle - \frac{1}{\sqrt{2}}.|111\rangle\right)$ <br> $|\psi_3\rangle = -\frac{1}{2}.|ppp\rangle + \frac{3}{2\sqrt{2}}.|111\rangle$ |
| Step 2.3. Application of $X^{\otimes 2} \otimes Id$ | $|\psi_4\rangle = -\frac{1}{2}.|ppp\rangle + \frac{3}{2\sqrt{2}}.|001\rangle$ |
|  | **Amplification** |
| Step 3.1. Application of $H^{\otimes 3}$ | $|\psi_5\rangle = -\frac{1}{2}.|000\rangle + \frac{3}{2\sqrt{2}}.|ppm\rangle$ |
| Step 3.2. Application of $X^{\otimes 3}$ | $|\psi_6\rangle = -\frac{1}{2}.|111\rangle - \frac{3}{2\sqrt{2}}.|ppm\rangle$ |
| Step 3.3. Application of $CZ(q_1, q_2; q_3)$ | $|\psi_7\rangle = \frac{1}{2}.|111\rangle - \frac{3}{2\sqrt{2}}.\left(|ppm\rangle + \frac{1}{\sqrt{2}}.|111\rangle\right)$ |

| | | |
|---|---|---|
| | | $\|\psi_7\rangle = -\frac{3}{2\sqrt{2}}.\|ppm\rangle - \frac{1}{4}.\|111\rangle$ |
| Step 3.4. Application of $X^{\otimes 3}$ | | $\|\psi_8\rangle = \frac{3}{2\sqrt{2}}.\|ppm\rangle - \frac{1}{4}.\|000\rangle$ |
| Step 3.5. Application of $H^{\otimes 3}$ | | $\|\psi_9\rangle = \frac{3}{2\sqrt{2}}.\|001\rangle - \frac{1}{4}.\|ppp\rangle$ |

## 5. Concluding remarks

In this paper we investigate a description of the Grover's algorithm using geometric considerations and a tensorial computations. The Grover's algorithm offers a promising way in solving OR problems where a solution has to be find under large scale search spaces with the promise of a quadratic speedup (Bourreau *et al.*, 2022). We introduce the Grover algorithm in a new way considering geometric consideration to illustrate how the successive quantum states are computed and how the planes are spanned by the different basis vector. The experiments have been achieved with the Qiskit library and meet the theoretical considerations.

## 7. Appendix: experiments with Qiskit and $n = 5$ qubits

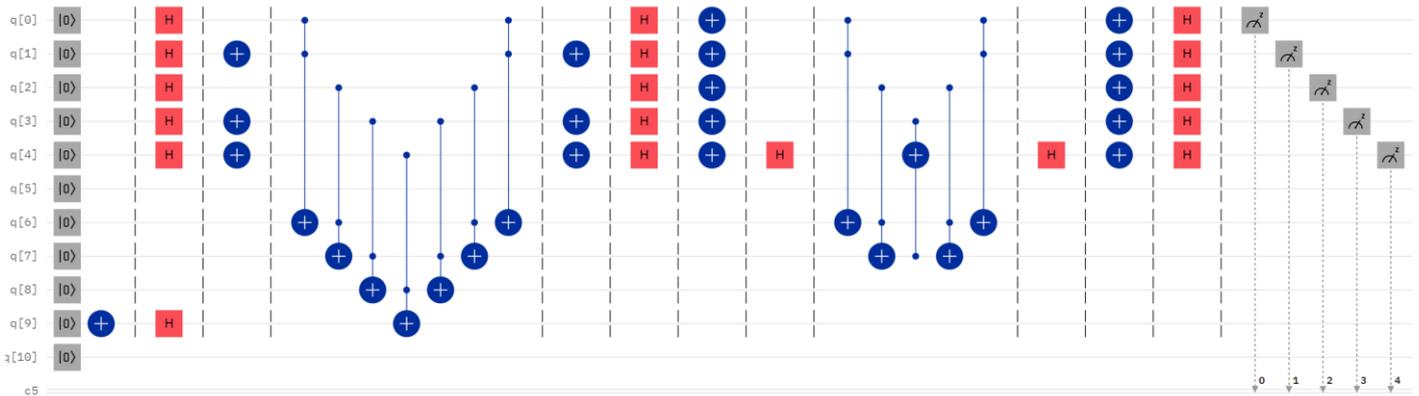

Figure A1. Numerical experiments with Qiskit ($n = 5$ and $\|\psi\rangle = \|10100\rangle$)

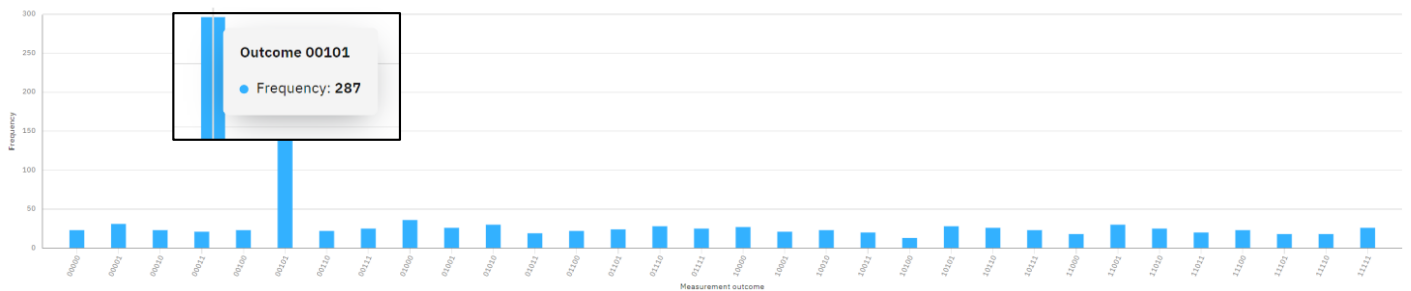

Figure A2. Numerical experiments with Qiskit ($n = 5$ and $|\psi\rangle = |10100\rangle$) and 1024 shots (one iteration): $P(|\psi\rangle = |r\rangle)$ estimated by $\frac{257}{1024} \simeq 25.1\%$

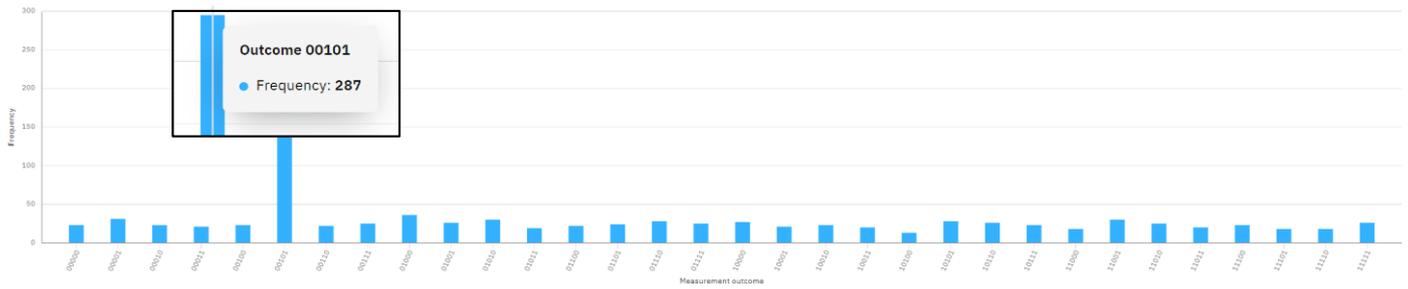

Figure A3. Numerical experiments with Qiskit ($n = 5$ and $|\psi\rangle = |10100\rangle$) and 1024 shots (seconds iteration): $P(|\psi\rangle = |r\rangle)$ estimated by $\frac{616}{1024} \simeq 60\%$